\documentstyle[prb,aps,twocolumn]{revtex}

\input{psfig}

\begin{document}
\twocolumn[
\hsize\textwidth\columnwidth\hsize\csname @twocolumnfalse\endcsname

\title{\hfill \\ Transport Properties of Highly Aligned Polymer
Light-Emitting-Diodes}
\author{A. Kambili} 
\address{Department of Physics, University of Bath, Claverton Down,
Bath BA2 7AY, UK, e-mail:pysak@bath.ac.uk}
\author{A. B. Walker}
\address{Department of Physics, University of Bath, Claverton Down,
Bath BA2 7AY, UK}
\date{\today} \maketitle
\begin{abstract}
We investigate hole transport in polymer light-emitting-diodes in
which the emissive layer is made of liquid-crystalline polymer
chains aligned perpendicular to the direction of transport.
Calculations of the current as a function of time via a random-walk
model show excellent qualitative agreement with experiments
conducted on electroluminescent polyfluorene demonstrating
non-dispersive hole transport. The current exhibits a constant
plateau as the charge carriers move with a time-independent drift
velocity, followed by a long tail when they reach the collecting
electrode. Variation of the parameters within the model allows the
investigation of the transition from non-dispersive to dispersive
transport in highly aligned polymers. It turns out that large
inter-chain hopping is required for non-dispersive hole transport
and that structural disorder obstructs the propagation of holes
through the polymer film. 
\end{abstract}
]
During the last few years there has been great development both in
the understanding of the underlying physics and the use of conjugated
polymers as the emissive layer in light-emitting-diodes \cite{1}.
Recently, there has been much interest in organic electroluminescent
devices which are based on aligned conjugated polymers that emit
polarized light \cite{2} since the latter could be used directly in
conventional liquid crystal displays \cite{3}. Time-of-flight
experiments in films of electroluminescent polyfluorene \cite{4}
have shown non-dispersive hole transport with high mobility and weak
field dependence, which indicates low energetic and positional
disorder. Moreover, it has been demonstrated \cite{5} that mesophase
alignment can enhance carrier mobilities for conjugated main-chain
liquid-crystalline polymers resulting in a more than one order of
magnitude increase in time-of-flight hole mobility normal to the
alignment direction. 

Previous theoretical studies have concentrated on the identification
of the field and temperature dependence of the carrier mobility in
disordered molecular materials, such as molecularly doped polymers,
small molecule glasses and conjugated polymers, in terms of the
disorder model \cite{6}. The latter assumes that the charge
transporting states, eg. dopant molecules, are energetically
localized, and the energy of these states is subject to a random
distribution introduced by disorder. Nevertheless, recent
experimental activity concerns the performance of devices made of a
layer of liquid-crystalline polymers, which are systems of high order
and orientation. In this paper, we present a simple model, based on
random walks, which studies charge transport through a film of highly
aligned polymers, and explains the mechanisms that enhance or hinder
current in the simplest possible way. The model simulates charge
carrier transport through an organic layer of liquid-crystalline
polymers oriented perpendicular to the direction of transport and
sandwiched between two electrodes, under the influence of an
externally applied electric field $E$. Charge carriers of one type
are injected in one of the electrodes, and by moving within the
polymer film due to the presence of the field $E$ they reach the
other electrode where they are discharged. We assume injection of
holes in accordance with the experiments in which it was not possible
to measure any current from electrons as the latter current is highly
dispersive. The term "dispersive" has been used in various contexts
and with different meanings \cite{7}. In this context dispersive
transport describes the situation where the mean velocity of the
carriers decreases with time. Our calculation of the current as a
function of time shows a constant current plateau followed by a
long tail, a behaviour which was observed in experiments on
electroluminescent polyfluorene and was linked to excellent purity
and chemical regularity of the material \cite{4,5}, and is typical
of Gaussian propagating packets. Investigation of the hopping
probabilities across the film of highly aligned polymers clearly
demonstrates that the non-dispersive character of hole transport is
retained only in the case of adequate inter-chain hopping along the
direction of transport, becoming dispersive otherwise. An additional
feature implemented in our model is structural order which is also
important for transport, since positional defects (voids) act as
traps, immobilizing the charge carriers within the polymer film and
resulting in dispersive hole current.

The polymer film is described by a two-dimensional discrete lattice,
with $x$ being the direction of transport and with periodic boundary
conditions along the $y$-direction. The polymer film is of thickness
$d=Da_{x}$ sites, where $D=100$ and the "lattice constant" is
$a_{x}=10 \AA$. The choice of $a_{x}$ was based on the fact that in
real devices liquid-crystalline polymer chains have a typical
inter-chain distance of about $12 \AA$ \cite{8}. We have considered
that the polymer chains are rigid rods, basing this approximation on
the extended "backbone" conjugation of most polymers that makes them
stiff and on bond vibrations being of high frequency and very low
amplitude. The chains are of length $L=10a_{y}$, where $a_{y}=1\AA$,
and are all oriented along the $y$-direction. The construction of
the polymer film is based on a random process. After choosing randomly
any of the equivalent sites of the lattice, we extend the chain
across the transverse direction only if a line of length $L$ along
$y$ is unoccupied. This prohibits both the occupation of the same
space by more than one chain and any cross-linking between different
chains. As a result, there is an upper limit ($\sim77 \%$) on the
density of polymer chains that can be deposited between the
electrodes which depends on the initial site chosen. The change in
this upper limit between different configurations is of the order of
$0.5\%$ or less, so that its effect on the resulting current is
negligible. The holes are injected into the polymer film at the anode
and they move towards the collecting electrode under the effect of an
external (positive) electric field $E$. The motion of the charge
carriers is due to hopping, which is simulated via a random-walk
process. Each carrier at a particular site hops to one of its
nearest-neighbouring sites on a polymer chain, with $p_{+x}$,
$p_{-x}$, being the probabilities for hopping along the
positive (right), negative (left) $x$-direction, respectively,
describing inter-chain hopping, and with $p_{+y}$, $p_{-y}$ being the
probabilities for hopping along the positive (up), negative (down)
$y$-direction, respectively, which account for intra-chain hopping.
Because of the presence of the positive field in the direction of
transport $p_{+x}$ is larger than $p_{-x}$, $p_{+y}$, and $p_{-y}$,
and the strength of the electric field is $|E|\propto p_{+x}-p_{-x}$.
Since the polymer chains are aligned normal to the direction of
transport the biased hopping probability indicates that inter-chain
hopping should be an important parameter. The drift of the carriers
under the external electric field results in a time-dependent current
$I(t)$. Each step of the random walk corresponds to one unit of time,
$\delta t=1$, in arbitrary units. The total current at each time $t$
is then given by \cite{9} 

\begin{equation}I(t)=-\frac{\delta}{\delta t}\sum_{x=1}^{d}\rho(x)+
\frac{1}{d}\frac{\delta}{\delta t}\sum_{x=1}^{d}x\rho(x)
\label{current}
\end{equation}
where $\rho(x)$ is the charge density, integrated over $y$. The first
term of equation \ref{current} gives the conduction current, whereas
the second term corresponds to the displacement current, for which
detailed investigation has shown that it is a small correction to the
conduction term.

\begin{figure}[h]
\centerline{\psfig{figure=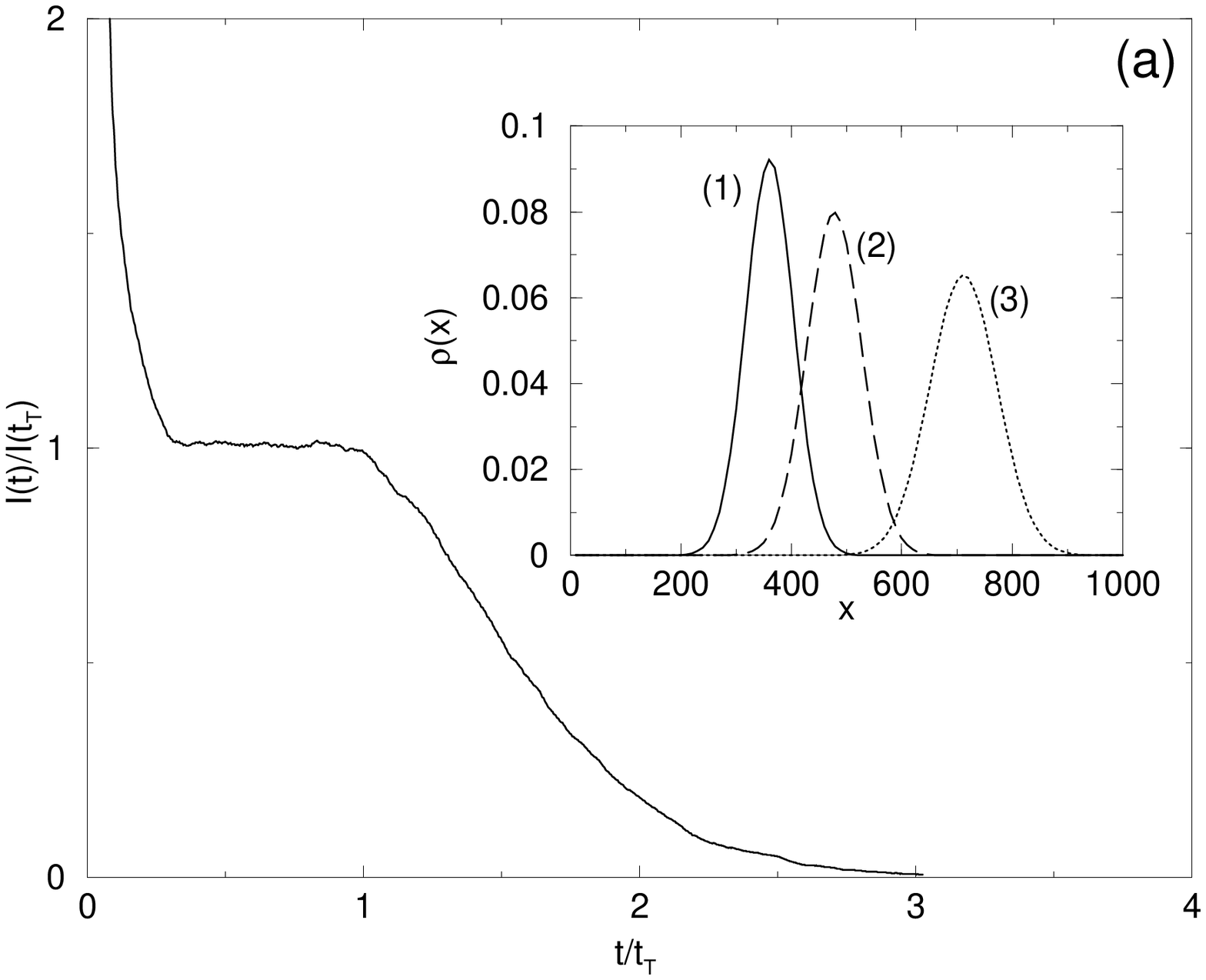,width=7cm}}
\centerline{\psfig{figure=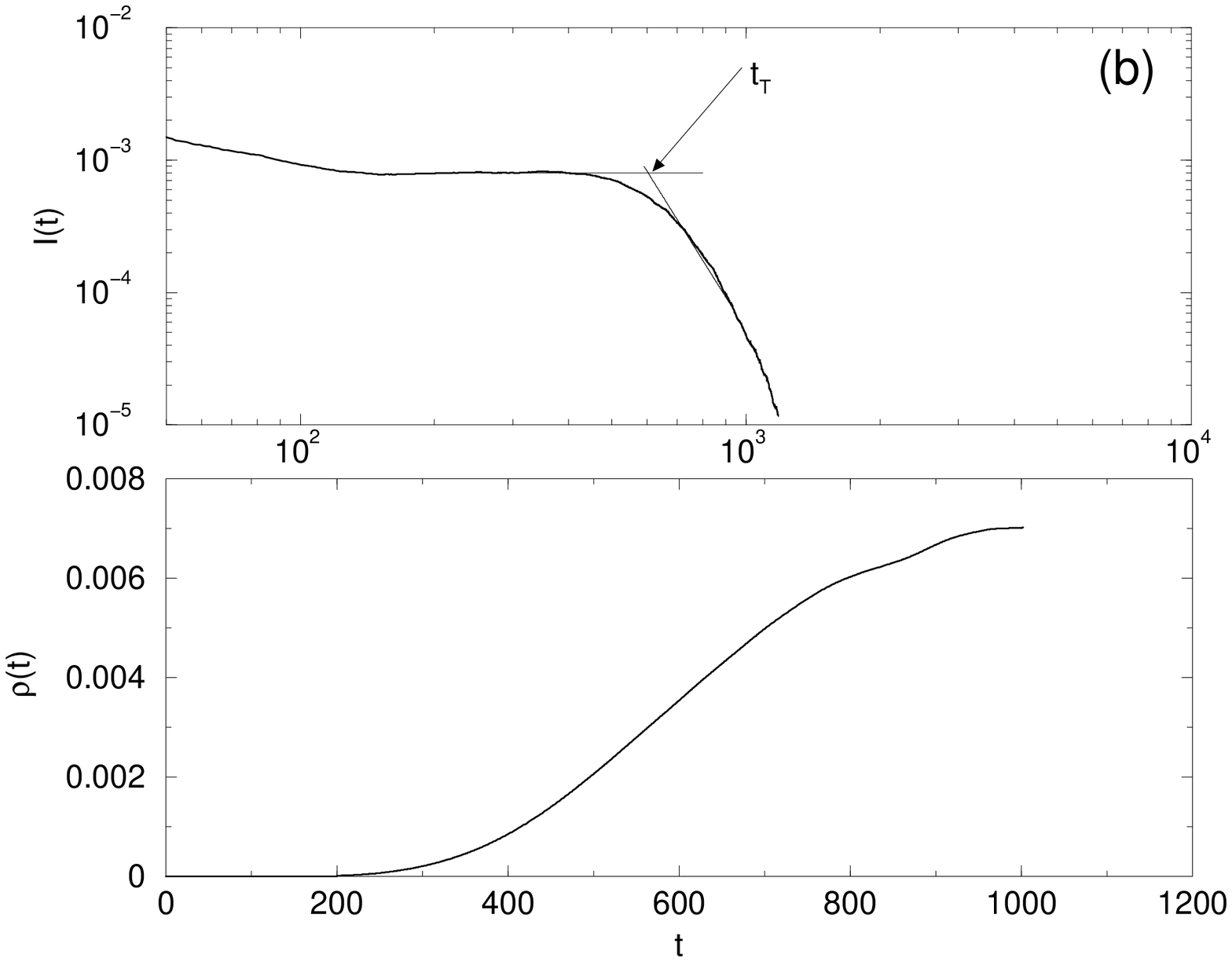,width=7.5cm}}
\caption{(a) Current versus time for hopping probabilities
$p_{+x}=0.4$, $p_{-x}=0.1$, $p_{+y}=0.25$, $p_{-y}=0.25$. In the
inset the charge density is calculated for three different times:
(1) $t/t_{T}=0.3$, (2) $t/t_{T}=0.4$, (3) $t/t_{T}=0.6$. (b) The
upper plot shows the double logarithmic representation of the current
with respect to time. The arrow indicates the transit time $t_{T}$
The lower graph shows the charge accumulated at a capacitor connected
at the collecting electrode at each time step. At both graphs the
axes are in arbitrary units.}
\label{1}
\end{figure}

Figure \ref{1}(a) shows the current transient from such simulations
obtained for hopping probabilities $p_{+x}=0.4$, $p_{-x}=0.1$,
$p_{+y}=0.25$ and $p_{-y}=0.25$. For small times the current shows a
spike, which becomes a constant plateau as time increases,
indicating that the injected carriers have the same mobility
$\mu$, as observed in the experiments \cite{4,5}. This plateau is
followed by a long tail when most of the charge carriers are
discharged at the second electrode. This shape of the current
transient is typical of Gaussian packets which propagate with time
\cite{10}. The latter has been verified in our simulations, as shown
in the inset of figure \ref{1}(a) where the charge density $\rho(x)$
is plotted for three different times. When the charge carriers are
injected at one of the electrodes, $\rho(x)$ is a $\delta$-function
centered at the electrode. When an electric field $E$ is applied, the
carriers move inside the film, and $\rho(x)$ is described by an
extended Gaussian distribution which propagates with time. The upper
graph of figure \ref{1}(b) shows the same current transient as in
\ref{1}(a) in a double logarithmic scale. The intersection point of
the asymptotes (indicated by the arrow) defines the transit time for
the arrival of the carriers at the collecting electrode. The transit
time is related to the carrier mobility $\mu$ and the external
electric field $E$ via the relation $\mu=d/(t_{T}E)$, and it is the
experimentally measured quantity from which the carrier mobility is
derived.

\begin{figure}[h]
\centerline{\psfig{figure=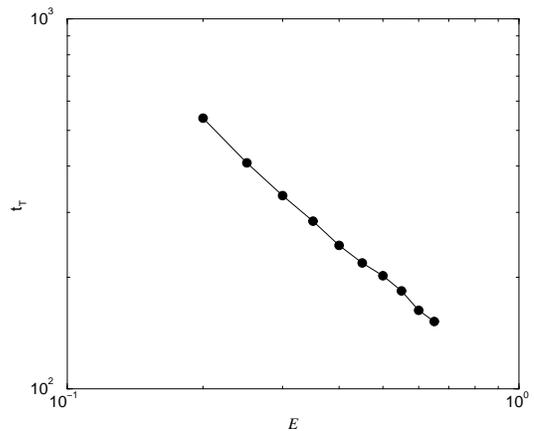,width=7cm}}
\caption{Double logarithmic representation of the transit time
$t_{T}$ versus the externally applied electric field
$|E|\propto p_{+x}-p_{-x}$, in arbitrary units.}
\label{2}
\end{figure}

An alternative way to measure $t_{T}$ experimentally is by connecting
the discharging electrode to a capacitor and measuring the charge that
accumulates there. The time at which the charge in the capacitor has
half of its maximum value must be equal to $t_{T}$. The lower graph of
figure \ref{1}(b) shows the results of such a numerical calculation.
Both techniques yield the same value of $t_{T}$, which is around
$t_{T}=600 \delta t$. In all the results presented in this paper we
have normalized the current transients to their value at the transit
time. Figure \ref{2} shows the transit time $t_{T}$ as a function of
the electric field $E$. Since $t_{T}$ decreases linearly as the
electric field becomes larger, we conclude that the mobility will have
a weak dependence on the electric field, as was seen in experiments
\cite{4,5}.

\begin{figure}[h]
\centerline{\psfig{figure=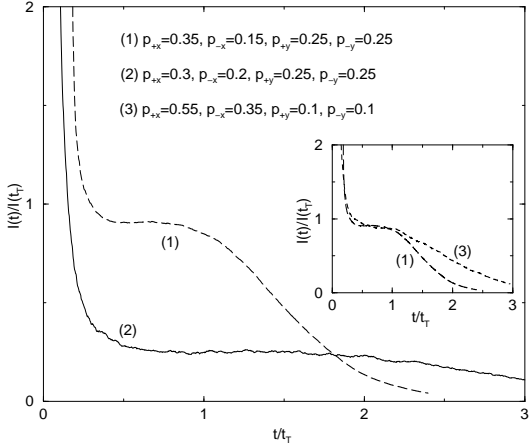,width=7cm}}
\caption{Current versus time for two different strengths of the
externally applied electric field $E$ (different hopping probabilities
across $x$). In the inset we have plotted the current versus time for
two different hopping probabilities across the $y$-direction.}
\label{3}
\end{figure}

So far we have demonstrated that a hopping model based on random
walks describes adequately hole transport within a polymer film of
highly aligned polymers, being in agreement with the experiments
which show non-dispersive hole currents. Our next aim is to point out
which are the parameters that affect transport, and when the current
changes from non-dispersive to dispersive. Figure \ref{3} shows the
current transients for different applied fields, thus for different
hopping probabilities in the direction of transport. Curve (1)
corresponds to $|E|\propto p_{+x}-p_{-x}=0.2$, which is smaller than
that of figure \ref{1}(a). The current plateau is retained and the
transport remains non-dispersive. However, if we reduce further the
electric field, as in curve (2), the plateau disappears and the
current decreases rapidly. Note that even though the plateau does
not exist anymore so that we cannot identify any transient time in
this case, we have divided the current of curve (2) with $I_{T}$ of
curve (1) to make a direct comparison between the two curves. 
The decrease of the current means that most of the charge remains
trapped within the polymer film and much fewer carriers make it to
the collecting electrode. If we keep the same strength of the electric
field, $E\propto 0.2$, but decrease $p_{+y}$ and $p_{-y}$, as is shown
in the inset of figure \ref{3} (curve (3)), the current remains
non-dispersive with a larger tail, thus, more holes will reach the
collecting electrode. We conclude from these results that in systems
of liquid-crystalline polymers inter-chain hopping (hopping along the
$x$-direction) is the essential mechanism for the transport of holes
(inter-chain hopping along $y$ is not permitted in our model).

\begin{figure}[h]
\centerline{\psfig{figure=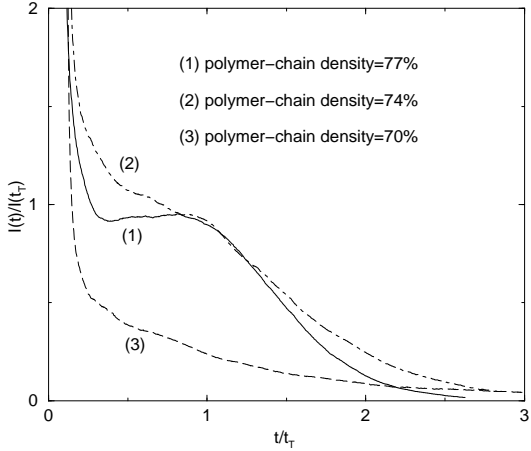,width=7cm}}
\caption{Current versus time for different densities of polymer
chains between the two electrodes ($p_{+x}=0.4$, $p_{-x}=0.1$,
$p_{+y}=0.25$, $p_{-y}=0.25$)}
\label{4}
\end{figure}

Figure \ref{4} shows the current transient for different densities of
polymer chains deposited in the emissive layer. If we allow fewer
polymer chains than the maximum allowed occupation density of the space
between the electrodes, the plateau is destroyed (curve (2)),
while upon further decrease the current becomes dispersive once again
(curve (3)). This is due to the presence of large gaps in the space
between the electrodes (voids), which act as traps for the carriers.
A threshold density of chains, dependent upon the chain
configuration, is found (68\% in this case) below which transport of
the carriers from one electrode to the other is totally prohibited.
In other words, positional disorder strongly affects transport and
a highly ordered polymer film is required for best device
performances.  

\begin{figure}[h]
\centerline{\psfig{figure=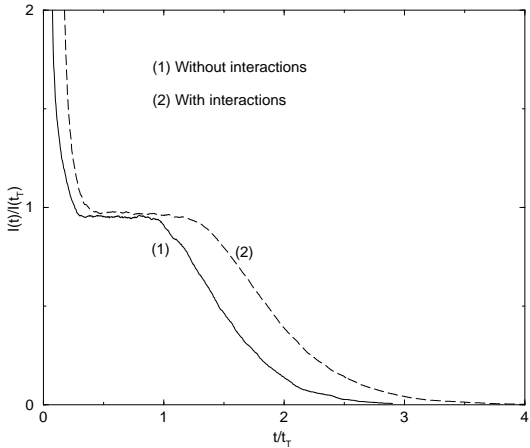,width=7cm}}
\caption{Current versus time with respect to the presence or absence
of on-site repulsive interactions between the charge carriers
($p_{+x}=0.4$, $p_{-x}=0.1$, $p_{+y}=0.25$, $p_{-y}=0.25$).}
\label{5}
\end{figure}

All the above results have been obtained without taking into account
the effect of repulsive interactions between the charge carriers.
Inclusion of the latter interaction can be incorporated within our
model by not allowing any two carriers to occupy the same site at
each time step. If all the neighbouring sites of a carrier are
occupied, it remains in the same position until one of these sites
becomes free, otherwise it gets trapped in this site. Figure \ref{5}
shows the current with respect to time in the absence (curve (1))
and the presence (curve(2)) of on-site repulsive interactions. In the
latter case the qualitative behaviour of the current does not change,
even though the plateau is larger and clearer and it takes more time
for the charge carriers to reach the second electrode and get discharged.

In summary, our random-walk model for the description of hole
transport in light-emitting-diodes whose active element is a polymer
film of highly aligned conjugated polymers, has added to the
understanding of the phenomenology of such devices. The
experimentally measured non-dispersive current transient has been
verified by the numerical simulations, which also show the constant
current plateau and the long tail, typical of Gaussian propagating
packets. Moreover, the model specifies the conditions under which
transport becomes dispersive, by stressing out the importance of
inter-chain hopping for transport in these liquid-crystalline polymers.
The effect of structural disorder is also discussed in connection
with better device performance, underlining the need of very clean
and ordered polymer films in order to avoid the presence of traps,
which will hinder transport and result in poor device performances.

The authors would like to thank Prof. A. M. Marshall and Prof D. D. C.
Bradley for their useful comments and remarks.

\end{document}